\documentclass[10pt,a4paper]{article}  
 \usepackage{geometry}               
 \usepackage{amsmath,mathrsfs,amssymb,graphicx,float,amsfonts}
 \usepackage{enumerate}              
 \usepackage{hyperref}               
 \usepackage{setspace}               
 \usepackage{lscape}
 \usepackage{empheq}                 

\begin{document}

\title{Thermodynamic Representations of Algebraically Modified FLRW Dynamics at the Apparent Horizon}
\author{Xiao Liu\\
\small School of Mathematics and Science, Hebei GEO University, Shijiazhuang 052161, Hebei, China\\
\small \texttt{liux@hgu.edu.cn}}
\date{}
\maketitle

\begin{abstract}
We investigate the thermodynamic description of a general class of algebraically modified Friedmann dynamics in which the physical matter obeys the standard conservation equation and the modification is encoded through a function \(g(\rho)\). By introducing a thermodynamic matter representation \(\chi(\rho)\), we construct the corresponding matter variables, heat flow, work term, and horizon entropy, and show that both the Clausius relation and the first-law balance can be consistently maintained. A continuous family of representations is then defined by \(\chi_\lambda(\rho)=(1-\lambda)\rho+\lambda g(\rho)\) with \(0\leq\lambda\leq1\), and the physical-matter and effective-fluid descriptions as the two endpoints. The associated thermodynamic quantities transform consistently between these limits. We further show that apparent departures from equilibrium can arise from mixing quantities belonging to different representations, rather than from a genuine non-equilibrium entropy-production mechanism. These results identify a continuous freedom of thermodynamic representation within this class of modified FLRW dynamics.
\end{abstract}

\section{Introduction}
\label{sec:intro}

The thermodynamic properties of gravitational horizons were first recognized in black hole physics, where the horizon entropy is proportional to its area and the temperature is determined by the surface gravity. Similar ideas have been extended to cosmological horizons, particularly to the apparent horizon of a FLRW universe. In this setting, the horizon dynamics can be formulated in terms of Hayward's unified first law \cite{haywardUnifiedFirstLaw1998},
\[
\mathrm{d}E_\mathrm{MS}=A\Psi+W\mathrm{d}V_A,
\]
which relates the variation of the Misner--Sharp energy to the energy-supply and work terms. Together with the Clausius relation and the Bekenstein--Hawking area entropy, this framework provides a thermodynamic description of the FLRW apparent horizon and its connection with the Friedmann equations \cite{caiFirstLawThermodynamics2005,caiUnifiedFirstLaw2007}.

The thermodynamic description is not unique once the cosmological dynamics is modified. In \(f(R)\) and related theories, non-equilibrium formulations with an entropy-production term can arise at the apparent horizon \cite{akbar2007ThermodynamicBehaviorField}, while an equilibrium description can also be obtained through a suitable redefinition of the effective matter sector \cite{bamba2010Equilibriumthermodynamicsmodified}. These results illustrate that the same underlying cosmological dynamics may admit different thermodynamic formulations, depending on how the modified contributions are assigned among the matter sector, the horizon entropy, and possible entropy-production terms. More recent analyses have emphasized this non-uniqueness explicitly \cite{pai2026Subtletiesnonequilibriumhorizona}. However, the additional geometric degrees of freedom in modified-gravity theories make the situation more involved: redefinitions of the matter sector are generally entangled with genuine geometric corrections to the horizon entropy and, in some formulations, with entropy-production terms. As a result, although different thermodynamic descriptions are known to exist, their transformation into one another is not easily isolated in a simple quantitative form. This motivates us to first consider a simpler class of modified cosmological dynamics in which the modification enters only algebraically through the matter density.

A particularly simple setting is provided by cosmological models in which the Friedmann equation is modified algebraically in the matter density while the standard conservation equation for the physical matter is retained. Cardassian cosmology is an early representative example, in which the usual linear dependence \(H^2\propto\rho\) is replaced by a nonlinear function \(H^2\propto g(\rho)\) \cite{freese2002CardassianExpansionModel,freese2005CardassianExpansionDark}. At the level of the effective background dynamics, loop quantum cosmology \cite{agulloLoopQuantumCosmology2016} provides another example of this structure. There, the modified contribution can be absorbed into an effective energy density and pressure, allowing the apparent-horizon first law to retain the standard area entropy \cite{liThermodynamicsLoopQuantum2009,zhangThermodynamicsNewModel2021}. A complementary description assigns the modification instead to the horizon entropy while keeping the physical matter variables unchanged, as illustrated by derivations of modified Friedmann equations from corrected entropy-area relations \cite{caiCorrectedEntropyAreaRelation2008}. These two types of description suggest natural endpoints between which a more general representation may be constructed.

For a general algebraically modified Friedmann dynamics, the relation between these thermodynamic descriptions has not, to our knowledge, been formulated as a continuous representation framework. Since the modification is encoded only through a function \(g(\rho)\) while the physical matter remains locally conserved, this class allows the representation change itself to be isolated from additional geometric degrees of freedom. We show that the equilibrium thermodynamic descriptions form a one-parameter family defined by
\[
\chi_{\lambda}(\rho)=(1-\lambda)\rho+\lambda g(\rho),\qquad 0\leq\lambda\leq1.
\]
The two endpoints correspond respectively to the physical-matter description with a corrected horizon entropy and the effective-fluid description with the standard area entropy. The associated thermodynamic quantities transform consistently along this family, so that \(\lambda\) labels different representations of the same cosmological dynamics rather than different cosmological models. This also makes clear that mixing quantities belonging to different representations can generate an apparent imbalance, but does not by itself imply a genuine non-equilibrium regime.

We use natural units \(c=G=\hbar=1\) throughout.

\section{Apparent-horizon thermodynamics in algebraically modified FLRW dynamics}
\label{sec:framework}

We consider a general class of algebraically modified Friedmann dynamics in a FLRW spacetime,
\[
\mathrm{d}s^2=h_{ab}\mathrm{d}x^a\mathrm{d}x^b+R^2\mathrm{d}\Omega^2,
\]
where \(h_{ab}=\mathrm{diag}(-1,a^2/(1-kr^2))\) is the metric of the two-dimensional subspace with coordinates \(x^0=t\) and \(x^1=r\), \(R=a(t)r\) is the areal radius, and \(k\) denotes the spatial curvature. The modification is assumed to enter only through an algebraic function of the physical matter density,
\begin{equation}
H^2+\frac{k}{a^2} = \frac{8\pi}{3}g(\rho),\qquad g(\rho)\in \mathcal{C}^1,
\label{eqFriedmann}
\end{equation}
while the physical matter obeys the standard conservation equation
\begin{equation}
\dot{\rho}+3H(\rho+P) = 0.
\label{eqConserv}
\end{equation}
We further impose \(g(0)=0\), thereby excluding an additive vacuum-energy contribution from the present framework. Differentiating Eq.~\eqref{eqFriedmann} and using Eq.~\eqref{eqConserv} gives the modified Raychaudhuri equation
\begin{equation}
\dot{H}-\frac{k}{a^2} = -4\pi g'(\rho)(\rho+P),
\label{eqRaych}
\end{equation}
where a prime denotes differentiation with respect to \(\rho\). Thus, the background modification is encoded in \(g(\rho)\), whereas \(g'(\rho)\) determines how the modified expansion responds to the evolution of the physical matter. 

The apparent horizon is defined as the marginal surface satisfying \(h^{ab}\partial_aR\partial_bR=0\). Denoting its areal radius by \(R_A\), this condition gives \cite{caiFirstLawThermodynamics2005}
\begin{equation}
R_A=\frac{1}{\sqrt{H^2+k/a^2}}.
\label{eqRA}
\end{equation}
For the expanding branch \(H>0\), differentiation with respect to cosmic time, together with Eq.~\eqref{eqRaych}, yields
\begin{equation}
\dot{R}_A = -HR_A^3\left(\dot{H}-\frac{k}{a^2}\right) = 4\pi HR_A^3g'(\rho)(\rho+P).
\label{eqDotRA1}
\end{equation}
This relation directly connects the evolution of the apparent horizon with the physical matter variables and will be used repeatedly below.

The Misner--Sharp energy enclosed within a sphere of radius \(R\) is \(E_\mathrm{MS}(R)=\frac{R}{2}\left(1-h^{ab}\partial_aR\partial_bR\right)\). On the apparent horizon it reduces to
\begin{equation}
E_\mathrm{MS}=\frac{R_A}{2}.
\label{eqMSenergy}
\end{equation}
This quantity is purely geometric. By contrast, the energy carried by the physical matter inside the apparent horizon is \(E=\rho V_A\), with \(V_A=4\pi R_A^3/3\). These two notions of energy need not coincide for a general \(g(\rho)\). For the physical matter, the work density is defined by
\begin{equation}
W = -\frac{1}{2}T^{ab}h_{ab} = \frac{\rho-P}{2}.
\label{eqWdens}
\end{equation}
where \(T^{ab}\) is the energy–momentum tensor of the cosmic fluid. The corresponding energy-supply one-form is defined by
\begin{align}
\Psi_a = T_a^{\ b}\partial_bR+W\partial_aR = \left(-\frac{1}{2}(\rho+P)HR,\,\,\frac{1}{2}(\rho+P)a\right).
\label{eqSupvec}
\end{align}
The work term \(W\mathrm{d}V_A\) accounts for the contribution associated with the change of the volume enclosed by the apparent horizon, while \(\Psi_a\) characterizes the matter energy flux. To obtain the heat flow across the dynamical horizon, \(\Psi_a\) is projected along the apparent-horizon trajectory. Following Hayward's trapping-horizon construction and its application to the FLRW apparent horizon \cite{caiUnifiedFirstLaw2007}, we parameterize this trajectory by cosmic time and introduce the tangent vector
\begin{equation}
\xi^a = \left(1,\,\,(\dot{R}_A-HR_A)/a\right).
\end{equation}
The heat flow through the apparent horizon during the time interval \(\mathrm{d}t\) is then
\begin{equation}
\delta Q = A\Psi_a\xi^a|_A\mathrm{d}t = 2\pi R_A^2(\rho+P)(\dot{R}_A-2HR_A)\mathrm{d}t,
\label{eqdefheatflow}
\end{equation}
where the negative-heat-out convention is adopted. 

The standard Bekenstein--Hawking entropy associated with the apparent horizon is
\begin{equation}
S_A=A/4=\pi R_A^2,
\label{eqSA}
\end{equation}
with \(A=4\pi R_A^2\). For the dynamical FLRW apparent horizon, the Hayward surface gravity is
\begin{equation}
\kappa=-\frac{1}{R_A}\left(1-\frac{\dot{R}_A}{2HR_A}\right).
\label{eqDefkappa}
\end{equation} 
Throughout this work we use the signed Hayward temperature
\begin{equation}
T_A= -\frac{\kappa}{2\pi}.
\label{eqDefTA}
\end{equation}
This choice retains the sign information carried by the surface gravity rather than replacing it by its absolute value. Since the thermodynamic representations considered below are formulated with a positive horizon temperature, we restrict the discussion to the sector \(T_A>0\). This restriction is a consistency condition of the present thermodynamic description and is not assumed to be a universal property of the FLRW apparent horizon.

\section{Thermodynamic representation of the apparent horizon}
\label{sec:representation}

We introduce a thermodynamic matter representation specified by a differentiable function \(\chi(\rho)\),  and define the represented energy density as
\begin{equation}
\rho_\chi = \chi(\rho),\qquad \chi(\rho)\in \mathcal{C}^1,
\label{eqdefrhochi}
\end{equation}
To describe this representation in the form of a locally conserved effective fluid, we define the corresponding pressure \(P_\chi\) through
\begin{equation}
\dot{\rho}_\chi+3H(\rho_\chi+P_\chi) = 0,
\label{eqConservchi}
\end{equation}
where \(\dot{\rho}_\chi=\chi'(\rho)\dot{\rho}\). Combining the physical matter conservation equation Eq.~\eqref{eqConserv} with this equation, we obtain
\begin{equation}
P_\chi = \chi'(\rho)(\rho+P)-\chi(\rho).
\label{eqdefPchi}
\end{equation} 
It follows that
\begin{equation}
\rho_\chi+P_\chi = \chi'(\rho)(\rho+P).
\label{eqPplusrhochi}
\end{equation}

Using these definitions, we derive the total energy and the work density associated with the \(\chi\) representation
\begin{align}
E_\chi &= \rho_\chi V_A = \chi(\rho)V_A,\\
W_\chi &= \frac{\rho_\chi-P_\chi}{2} = \chi(\rho)-\frac{1}{2}\chi'(\rho)(\rho+P).
\label{eqWdenschi}
\end{align}
From Eq.~\eqref{eqSupvec} and Eq.~\eqref{eqPplusrhochi}, the energy-supply one-form in the \(\chi\) representation becomes
\begin{equation}
\Psi_a^{(\chi)} = \chi'(\rho)\Psi_a.
\label{eqSupvecchi}
\end{equation}
Here, the representation label is placed in parentheses as superscripts to avoid confusion with spacetime indices. Substituting \(\Psi_a^{(\chi)}\) into the definition Eq.~\eqref{eqdefheatflow}, the heat flow through the apparent horizon in the \(\chi\) representation is
\begin{equation}
\delta Q_\chi = A \Psi_a^{(\chi)}\xi^a|_A\mathrm{d}t=\chi'(\rho)\delta Q.
\label{eqHeatflowchi}
\end{equation}

To determine the entropy associated with the \(\chi\) representation, we first relate the physical-matter heat flow \(\delta Q\) to the geometrically defined Hayward temperature and the standard area entropy.
\begin{equation}
-T_A\mathrm{d}S_A = R_A\dot{R}_A\kappa\mathrm{d}t.
\end{equation}
Here we have used \(T_A=-\kappa/2\pi\) and \(\mathrm{d}S_A=2\pi R_A\mathrm{d}R_A\). From Eq.~\eqref{eqDotRA1} and Eq.~\eqref{eqDefkappa}, the right-hand side of the above equation is
\begin{align}
R_A\dot{R}_A\kappa\mathrm{d}t &= g'(\rho)4\pi HR_A^4(\rho+P)\kappa\notag\\
                              &= g'(\rho)2\pi R_A^2(\rho+P)(\dot{R}_A-2HR_A)\mathrm{d}t\notag\\
                              &= g'(\rho)\delta Q.
\end{align}
Consequently,
\begin{equation}
-T_A\mathrm{d}S_A = g'(\rho)\delta Q.
\label{eqClausiuseff}
\end{equation}

Combining Eq.~\eqref{eqHeatflowchi} with Eq.~\eqref{eqClausiuseff}, we obtain
\begin{equation}
\delta Q_\chi = -\frac{\chi'(\rho)}{g'(\rho)}T_A\mathrm{d}S_A.
\label{eqClausiuschi}
\end{equation}
We now require the \(\chi\) representation to retain the equilibrium Clausius form
\begin{equation}
\delta Q_\chi = -T_A\mathrm{d}S_\chi.
\end{equation}
With \(T_A\) kept fixed, comparison with Eq.~\eqref{eqClausiuschi} defines
\begin{equation}
\mathrm{d}S_\chi \equiv \frac{\chi'(\rho)}{g'(\rho)}\mathrm{d}S_A.
\label{eqdSchi}
\end{equation}
Eq.~\eqref{eqClausiuschi} does not uniquely determine how the factor \(\chi'(\rho)/g'(\rho)\) should be assigned between the temperature and entropy sectors. In the present work, we keep \(T_A\) fixed as the geometrically defined Hayward temperature and absorb this factor into the horizon entropy. Since \(\mathrm{d}S_A=2\pi R_A\mathrm{d}R_A\), Eq.~\eqref{eqdSchi} becomes
\begin{equation}
\mathrm{d}S_\chi = 2\pi R_A \frac{\chi'(\rho)}{g'(\rho)}\mathrm{d}R_A.
\label{eqdSchiA}
\end{equation}
From Eq.~\eqref{eqFriedmann} and Eq.~\eqref{eqRA}, we have
\begin{equation}
R_A^{-2}=\frac{8\pi}{3}g(\rho).
\label{eqRAinversesquare}
\end{equation}
On any branch where \(g'(\rho)\neq0\), Eq.~\eqref{eqRAinversesquare} locally defines \(\rho\) as a function of \(R_A\). Hence the coefficient \(\chi'(\rho)/g'(\rho)\) in Eq.~\eqref{eqdSchiA} may be regarded locally as a function of \(R_A\), so that \(\mathrm{d}S_\chi=F(R_A)\mathrm{d}R_A\) is integrable. The entropy \(S_\chi(R_A)\) is therefore determined up to an additive constant on each such branch.

Having established the Clausius relation in the \(\chi\) representation, we next verify that the corresponding energy and work variables satisfy the first-law balance on the apparent horizon. The differential of the total energy in this representation is
\begin{equation}
\mathrm{d}E_\chi = \mathrm{d}\left(\rho_\chi V_A\right)
= \dot{\rho}_\chi V_A\mathrm{d}t+\rho_\chi\dot{V}_A\mathrm{d}t.
\end{equation}
Using the conservation equation Eq.~\eqref{eqConservchi} and \(V_A=4\pi R_A^3/3\), we obtain
\begin{equation}
\rho_\chi\dot{V}_A\mathrm{d}t = 4\pi R_A^2\dot{R}_A\rho_\chi\mathrm{d}t,\qquad
\dot{\rho}_\chi V_A\mathrm{d}t = -4\pi HR_A^3\left(\rho_\chi+P_\chi\right)\mathrm{d}t.
\end{equation}
Then
\begin{align}
\mathrm{d}E_\chi &= 4\pi R_A^2\left(\dot{R}_A\rho_\chi-HR_A(\rho_\chi+P_\chi)\right)\mathrm{d}t\notag\\
                 &= 4\pi R_A^2\left(\dot{R}_A\frac{\rho_\chi+P_\chi}{2}-HR_A(\rho_\chi+P_\chi)+\dot{R}_A\frac{\rho_\chi-P_\chi}{2}\right)\mathrm{d}t\notag\\
                 &= 2\pi R_A^2\left(\rho_\chi+P_\chi\right)(\dot{R}_A-2HR_A)\mathrm{d}t+\frac{\rho_\chi-P_\chi}{2}4\pi R_A^2\dot{R}_A\mathrm{d}t.
\label{eqEnergydiff}
\end{align}
Comparing the above expression with Eq.~\eqref{eqWdenschi} and Eq.~\eqref{eqHeatflowchi},  the two terms in Eq.~\eqref{eqEnergydiff} are identified respectively as \(\delta Q_\chi\) and \(W_\chi\mathrm{d}V_A\). Hence
\begin{equation}
\mathrm{d}E_\chi = \delta Q_\chi+W_\chi\mathrm{d}V_A.
\end{equation}
Therefore, both the Clausius relation and the first-law balance are consistently closed within the same \(\chi\) representation. 
\section{Continuous thermodynamic representations}
\label{sec:consis}

On each branch where the entropy construction above is well defined, the equilibrium thermodynamic description of the modified FLRW apparent horizon is not unique. From the above calculations, it is natural to distinguish different representations by how much of the modified background dynamics is absorbed into the heat-flow sector (which also represents the matter sector). As shown in Eq.~\eqref{eqHeatflowchi}, the factor \(\chi'(\rho)\) controls how much of the modification is assigned to the matter/heat-flow sector. We therefore introduce a continuous variable \(\lambda\) satisfying
\begin{equation}
\chi'_\lambda(\rho) = (1-\lambda)+\lambda g'(\rho),\qquad0\leq\lambda\leq1,
\label{eqdefpara}
\end{equation}
to parameterize the thermodynamic matter representations.  Here \(\lambda\) parameterizes the allocation of the dynamical correction to the matter/heat-flow sector. This allocation fixes \(\chi_\lambda(\rho)\) up to an additive constant, which corresponds to a vacuum-like shift of the energy density. Since we consider algebraically modified Friedmann dynamics  satisfying \(g(0)=0\), this freedom can be fixed by a reference normalization \(\chi_\lambda(0)=0\). After this normalization, integrating Eq.~\eqref{eqdefpara} gives
\begin{equation}
\chi_\lambda(\rho) = (1-\lambda)\rho+\lambda g(\rho).
\label{eqparaint}
\end{equation} 

Among the general representations, two particular choices provide natural reference descriptions. When \(\lambda=0\), we have 
\[
\chi_0(\rho) = \rho.
\] 
This is the physical-matter representation \cite{caiCorrectedEntropyAreaRelation2008}. In this representation, all modifications relative to the standard Friedmann thermodynamic description are assigned to the entropy sector. To retain an equilibrium thermodynamic description in which the Clausius relation and the first law are maintained, a corrected entropy satisfying
\begin{equation}
\mathrm{d}S\equiv \mathrm{d}S_{\chi_0}=\frac{1}{g'(\rho)}\mathrm{d}S_A,
\label{eqdSmatter}
\end{equation}
has to be introduced. Then
\begin{equation}
\delta Q = -T_A\mathrm{d}S,\qquad \mathrm{d}E = \delta Q+W\mathrm{d}V_A.
\end{equation}
Here we use the convention that no representation label is attached when a thermodynamic quantity is defined by the physical matter. For example, \(\Psi^{(\chi_0)}_a\equiv\Psi_a\), \(\delta Q_{\chi_0}\equiv\delta Q\), and similarly for the remaining quantities. This convention is consistent with the discussion in previous sections.

When \(\lambda=1\), we have
\[
\chi_1(\rho) = g(\rho).
\]
This is the effective-fluid representation \cite{liThermodynamicsLoopQuantum2009, zhangThermodynamicsNewModel2021}, in which all modifications relative to the standard Friedmann thermodynamic description are absorbed into the matter sector. From Eq.~\eqref{eqdSchi}, in this case, we have
\begin{equation}
\mathrm{d}S_{\chi_1}=\mathrm{d}S_A.
\label{eqdSphymatter}
\end{equation} 
The corresponding effective-fluid variables are therefore
\begin{align}
\rho_{eff}\equiv\rho_{\chi_1}=\chi_1(\rho),\qquad &P_{eff}\equiv P_{\chi_1}=g'(\rho)(\rho+P)-g(\rho),\notag\\
E_{eff}\equiv E_{\chi_1}=\rho_{eff}V_A,\qquad &W_{eff}\equiv W_{\chi_1}=(\rho_{eff}-P_{eff})/2,\notag\\
\Psi^{(eff)}_a\equiv \Psi^{(\chi_1)}_a=g'(\rho)\Psi_a,\qquad &\delta Q_{eff}\equiv \delta Q_{\chi_1}=g'(\rho)\delta Q.
\end{align}
A direct result of the above definitions is
\begin{equation}
E_{eff} = \rho_{eff}V_A = g(\rho)V_A = \frac{3}{8\pi R_A^2}\frac{4\pi R_A^3}{3} = \frac{R_A}{2} = E_\mathrm{MS},
\label{eqMSeff}
\end{equation}
where Eq.~\eqref{eqRAinversesquare} has been used. Thus, at the effective-fluid endpoint, the thermodynamic total energy coincides exactly with the geometrically defined Misner--Sharp energy. The Clausius relation and the first law can be written as
\begin{equation}
\delta Q_{eff} = -T_A\mathrm{d}S_A,\qquad \mathrm{d}E_\mathrm{MS} = \delta Q_{eff}+W_{eff}\mathrm{d}V_A,
\end{equation}

Furthermore, combining Eqs.~\eqref{eqdefrhochi}-\eqref{eqHeatflowchi} and Eq.~\eqref{eqdSchi} with the definitions in the 
parameterized representations, we obtain the following sequence of relations:
\begin{empheq}[box={\setlength{\fboxsep}{10pt}\fbox}]{align}
\rho_{\chi_\lambda} &= (1-\lambda)\rho+\lambda\rho_{eff}\notag\\
P_{\chi_\lambda} &= (1-\lambda)P+\lambda P_{eff}\notag\\
E_{\chi_\lambda} &= (1-\lambda)E+\lambda E_\mathrm{MS}\notag\\
W_{\chi_\lambda} &= (1-\lambda)W+\lambda W_{eff}\notag\\
\Psi^{(\chi_\lambda)}_a &= (1-\lambda)\Psi_a+\lambda\Psi^{(eff)}_a\notag\\
\delta Q_{\chi_\lambda} &= (1-\lambda)\delta Q+\lambda\delta Q_{eff}\notag\\
\mathrm{d}S_{\chi_\lambda} &= (1-\lambda)\mathrm{d}S+\lambda\mathrm{d}S_A\notag
\end{empheq}
These relations show that the interpolation introduced through \(\chi_\lambda\) propagates consistently through the entire thermodynamic description. The parameter \(\lambda\) therefore does not merely redistribute the modified contribution in the matter sector; it simultaneously determines the corresponding energy density, pressure, total energy, work density, energy-supply one-form, heat flow, and horizon entropy. The seven quantities are not independently interpolated, but are generated by the same representation map \(\chi_\lambda\). Consequently, each value of \(\lambda\) defines a closed thermodynamic representation in which the Clausius relation and the first-law balance retain the same form. The physical-matter and effective-fluid descriptions are therefore not isolated constructions, but the two limiting members of this continuous family.

For a fixed \(g(\rho)\), changing \(\lambda\) therefore changes only the thermodynamic representation, not the underlying cosmological dynamics. The parameter \(\lambda\) should accordingly be regarded as a representation label rather than as a new physical degree of freedom. What is physically important is not the choice of \(\lambda\) itself, but the internal consistency of the quantities used within a given representation. This point becomes particularly transparent when quantities belonging to different representations are mixed.

For example, as shown in Eq.~\eqref{eqdSphymatter} and Eq.~\eqref{eqMSeff}, the original geometric definition of the  standard Bekenstein--Hawking area entropy and the Misner--Sharp energy are incorporated in the effective-fluid representation, in which all corrections are absorbed into the matter sector. If we wish to describe the apparent horizon using the classical quantities \(E_\mathrm{MS}\), \(S_A\) together with the physical matter variables \(\rho\), \(P\), \(W\), \(\Psi_a\) and  \(\delta Q\), then from Eq.~\eqref{eqClausiuseff} we will conclude that \(-T_A\mathrm{d}S_A = g'(\rho)\delta Q\), which means that the Clausius relation does not hold in this mixed description. If we further calculate the differential of the Misner--Sharp energy in this case, setting \(\lambda=1\) in Eq.~\eqref{eqEnergydiff} gives
\begin{align}
\mathrm{d}E_\mathrm{MS} &= 2\pi R_A^2\left(\rho_{eff}+P_{eff}\right)(\dot{R}_A-2HR_A)\mathrm{d}t+\frac{\rho_{eff}-P_{eff}}{2}4\pi R_A^2\dot{R}_A\mathrm{d}t. \notag\\
                 &= g'(\rho)\delta Q + W_{eff}\mathrm{d}V_A.
\end{align}
Subtracting the heat term \(-T_A\mathrm{d}S_A\) and the physical-matter work term \(W\mathrm{d}V_A\) from \(\mathrm{d}E_\mathrm{MS}\), we obtain
\begin{equation}
\mathrm{d}E_\mathrm{MS}+T_A\mathrm{d}S_A-W\mathrm{d}V_A = (W_{eff}-W)\mathrm{d}V_A.
\end{equation}  
This represents an apparent departure from the equilibrium thermodynamic balance, but the departure in this example arises entirely from mixing quantities belonging to different thermodynamic representations (energy/entropy at the endpoint \(\lambda=1\)  with heat/work at the endpoint \(\lambda=0\)). The resulting imbalance is therefore a consequence of representation mismatch rather than evidence for genuine non-equilibrium entropy production.

\section{Conclusion}
\label{sec:discuss}

For a general class of algebraically modified Friedmann dynamics that preserve the standard matter conservation equation, we have constructed a continuous family of equilibrium thermodynamic representations of the apparent horizon. The family is parameterized by \(\lambda\in[0,1]\) through
\[
\chi_\lambda(\rho)=(1-\lambda)\rho+\lambda g(\rho),
\]
where \(\lambda=0\) corresponds to the physical-matter representation with a corrected horizon entropy, while \(\lambda=1\) corresponds to the effective-fluid representation with the standard Bekenstein--Hawking area entropy.

Within this family, the energy density, pressure, total energy, work density, energy-supply one-form, heat flow, and horizon entropy are transformed consistently and are connected linearly between the two limiting representations. The parameter \(\lambda\) therefore labels different thermodynamic descriptions of the same cosmological dynamics rather than different cosmological models. Each value of \(\lambda\) defines an internally closed representation in which the Clausius relation and the first-law balance retain their equilibrium form.

We have also shown that apparent departures from equilibrium can arise when thermodynamic quantities belonging to different representations are combined. Such an imbalance reflects a mismatch of thermodynamic representation rather than, by itself, a genuine non-equilibrium entropy-production mechanism.

The present analysis is restricted to algebraically modified Friedmann dynamics with the standard matter conservation equation and does not directly extend to modified-gravity theories with independent geometric dynamical degrees of freedom. Within this scope, the results indicate that part of the apparent ambiguity in horizon thermodynamics can be understood as a freedom of thermodynamic representation rather than as a change in the underlying cosmological dynamics.

\bibliographystyle{unsrt}
\bibliography{references}

@misc{agulloLoopQuantumCosmology2016,
  title = {Loop {{Quantum Cosmology}}: {{A}} Brief Review},
  shorttitle = {Loop {{Quantum Cosmology}}},
  author = {Agullo, Ivan and Singh, Parampreet},
  year = 2016,
  month = dec,
  number = {arXiv:1612.01236},
  eprint = {1612.01236},
  primaryclass = {gr-qc},
  publisher = {arXiv},
  doi = {10.48550/arXiv.1612.01236},
  urldate = {2026-08-20},
  archiveprefix = {arXiv},
  langid = {english}
}

@article{akbar2007ThermodynamicBehaviorField,
  title = {Thermodynamic {{Behavior}} of {{Field Equations}} for f({{R}}) {{Gravity}}},
  author = {Akbar, M. and Cai, Rong-Gen},
  year = 2007,
  month = may,
  journal = {Physics Letters B},
  volume = {648},
  number = {2-3},
  eprint = {gr-qc/0612089},
  pages = {243--248},
  issn = {03702693},
  doi = {10.1016/j.physletb.2007.03.005},
  urldate = {2026-11-05},
  archiveprefix = {arXiv}
}

@article{bamba2010Equilibriumthermodynamicsmodified,
  title = {Equilibrium Thermodynamics in Modified Gravitational Theories},
  author = {Bamba, Kazuharu and Geng, Chao-Qiang and Tsujikawa, Shinji},
  year = 2010,
  month = apr,
  journal = {Physics Letters B},
  volume = {688},
  number = {1},
  eprint = {0909.2159},
  primaryclass = {gr-qc},
  pages = {101--109},
  issn = {03702693},
  doi = {10.1016/j.physletb.2010.03.070},
  urldate = {2026-10-28},
  archiveprefix = {arXiv}
}

@article{caiCorrectedEntropyAreaRelation2008,
  title = {Corrected {{Entropy-Area Relation}} and {{Modified Friedmann Equations}}},
  author = {Cai, Rong-Gen and Cao, Li-Ming and Hu, Ya-Peng},
  year = 2008,
  month = aug,
  journal = {Journal of High Energy Physics},
  volume = {2008},
  number = {08},
  eprint = {0807.1232},
  primaryclass = {hep-th},
  pages = {090--090},
  issn = {1029-8479},
  doi = {10.1088/1126-6708/2008/08/090},
  urldate = {2026-08-20},
  archiveprefix = {arXiv},
  langid = {english}
}

@article{caiFirstLawThermodynamics2005,
  title = {First {{Law}} of {{Thermodynamics}} and {{Friedmann Equations}} of {{Friedmann-Robertson-Walker Universe}}},
  author = {Cai, Rong-Gen and Kim, Sang Pyo},
  year = 2005,
  month = feb,
  journal = {Journal of High Energy Physics},
  volume = {2005},
  number = {02},
  eprint = {hep-th/0501055},
  pages = {050--050},
  issn = {1029-8479},
  doi = {10.1088/1126-6708/2005/02/050},
  urldate = {2026-08-20},
  archiveprefix = {arXiv},
  langid = {english}
}

@article{caiUnifiedFirstLaw2007,
  title = {Unified {{First Law}} and {{Thermodynamics}} of {{Apparent Horizon}} in {{FRW Universe}}},
  author = {Cai, Rong-Gen and Cao, Li-Ming},
  year = 2007,
  month = mar,
  journal = {Physical Review D},
  volume = {75},
  number = {6},
  eprint = {gr-qc/0611071},
  pages = {064008},
  issn = {1550-7998, 1550-2368},
  doi = {10.1103/PhysRevD.75.064008},
  urldate = {2026-08-20},
  archiveprefix = {arXiv},
  langid = {english}
}

@article{freese2002CardassianExpansionModel,
  title = {Cardassian {{Expansion}}: A {{Model}} in Which the {{Universe}} Is {{Flat}}, {{Matter Dominated}}, and {{Accelerating}}},
  shorttitle = {Cardassian {{Expansion}}},
  author = {Freese, Katherine and Lewis, Matthew},
  year = 2002,
  month = jul,
  journal = {Physics Letters B},
  volume = {540},
  number = {1-2},
  eprint = {astro-ph/0201229},
  pages = {1--8},
  issn = {03702693},
  doi = {10.1016/S0370-2693(02)02122-6},
  urldate = {2026-10-28},
  archiveprefix = {arXiv}
}

@article{freese2005CardassianExpansionDark,
  title = {Cardassian {{Expansion}}: {{Dark Energy Density}} from {{Modified Friedmann Equations}}},
  shorttitle = {Cardassian {{Expansion}}},
  author = {Freese, Katherine},
  year = 2005,
  month = may,
  journal = {New Astronomy Reviews},
  volume = {49},
  number = {2-6},
  eprint = {astro-ph/0501675},
  pages = {103--109},
  issn = {13876473},
  doi = {10.1016/j.newar.2005.01.007},
  urldate = {2026-10-28},
  archiveprefix = {arXiv}
}

@article{haywardUnifiedFirstLaw1998,
  title = {Unified First Law of Black-Hole Dynamics and Relativistic Thermodynamics},
  author = {Hayward, Sean A.},
  year = 1998,
  month = oct,
  journal = {Classical and Quantum Gravity},
  volume = {15},
  number = {10},
  eprint = {gr-qc/9710089},
  pages = {3147--3162},
  issn = {0264-9381, 1361-6382},
  doi = {10.1088/0264-9381/15/10/017},
  urldate = {2026-08-20},
  archiveprefix = {arXiv},
  langid = {english}
}

@article{liThermodynamicsLoopQuantum2009,
  title = {Thermodynamics in {{Loop Quantum Cosmology}}},
  author = {Li, Li-Fang and Zhu, Jian-Yang},
  year = 2009,
  month = jan,
  journal = {Advances in High Energy Physics},
  volume = {2009},
  number = {1},
  eprint = {0812.3544},
  primaryclass = {gr-qc},
  pages = {905705},
  issn = {1687-7357, 1687-7365},
  doi = {10.1155/2009/905705},
  urldate = {2026-08-20},
  archiveprefix = {arXiv},
  langid = {english}
}

@misc{pai2026Subtletiesnonequilibriumhorizona,
  title = {Subtleties in Non-Equilibrium Horizon Thermodynamics of Modified Gravity Theories},
  author = {Pai, Vishnu A. and Namboothiri, Vishnu S. and Mathew, Titus K.},
  year = 2026,
  month = apr,
  number = {arXiv:2604.04731},
  eprint = {2604.04731},
  primaryclass = {gr-qc},
  publisher = {arXiv},
  doi = {10.48550/arXiv.2604.04731},
  urldate = {2026-11-03},
  archiveprefix = {arXiv}
}

@article{zhangThermodynamicsNewModel2021,
  title = {Thermodynamics in New Model of Loop Quantum Cosmology},
  author = {Zhang, Xiangdong},
  year = 2021,
  month = feb,
  journal = {The European Physical Journal C},
  volume = {81},
  number = {2},
  eprint = {2111.05660},
  primaryclass = {gr-qc},
  pages = {117},
  issn = {1434-6044, 1434-6052},
  doi = {10.1140/epjc/s10052-021-08922-2},
  urldate = {2026-08-20},
  archiveprefix = {arXiv},
  langid = {english}
}

\end{document}